\documentclass[aps,preprintnumbers,eqsecnum,amsmath,amssymb,nofootinbib]{revtex4}  
\usepackage{graphicx}
\usepackage{bm}
\usepackage{epsfig}
\usepackage{color}
\usepackage{float}

\begin{document}
\title{Form factors for $B\to j_1j_2$ decays into two currents in QCD}
\author{Mikhail A.~Ivanov$^a$, Dmitri Melikhov$^{a,b,c}$, and Silvano Simula$^d$}
\affiliation{
$^a$Joint Institute for Nuclear Research, Dubna, 141980, Russia\\
$^b$D.~V.~Skobeltsyn Institute of Nuclear Physics, M.~V.~Lomonosov Moscow State University, 119991, Moscow, Russia\\
$^c$Faculty of Physics, University of Vienna, Boltzmanngasse 5, A-1090 Vienna, Austria\\
$^d$INFN, Roma Tre, Via della Vasca Navale 84, I-00146, Rome, Italy
}
\date{\today}
\begin{abstract}
We discuss the general properties of the $B$-meson decay form factors $F(q^2,q'^2)$, 
describing $B$-meson decays induced by two currents, e.g., 
$B$-decays into four leptons in the final state. We study the analytic properties of these 
complicated objects and identify those regions 
of $q^2$ and $q'^2$, where perturbative QCD can be applied for obtaining predictions for 
the physical form factors. 
\end{abstract}
\maketitle
\section{Introduction}
\label{Sec_introduction}
The goal of this paper is to study the general properties of the $B$-meson form factors into two currents.
Such quantities emerge, e.g., in the $B$-meson decays into four leptons in the final state; the latter reactions 
are being studied experimentally \cite{exp1,exp2,exp3,exp4}, thus requiring a proper theoretical understanding 
of the $B$-meson form factors into two currents. By now, there have been only few papers \cite{sehgal,nikitin}, 
where $B$-decays into two lepton pairs have been studied theoretically, using a rather naive model for the 
$B$-decay form factors. To obtain reliable predictions for four-lepton $B$-decays one needs a detailed understanding 
of the appropriate form factors in the timelike region. In particular, one needs to study the analytic properties 
of the latter in broad ranges of momentum transfers, including 
the timelike regions of four-lepton decays, in order to understand where model-independent predictions for such form 
factors directly from QCD may be obtained. The knowledge of rigorous properties of such form factors from field theory 
allows one to obtain constraints to be used in building realistic phenomenological models for the amplitudes 
of various four-lepton decays. 

In the past, theoretical analyses focused on a family of similar reactions, namely, 
the $B\to\gamma l^+l^-$ and $B\to \gamma l\nu$ decays (see, e.g., \cite{kruger,korchemsky,mn,kmn2018,ivanov}); these processes are 
described by the same form factors as four-lepton $B$-decays, but evaluated for a zero value of one of the momenta squared. 
The corresponding form factors depend on one variable $q^2$, $q$ being the momentum of the weak current. 
Various versions of light-cone QCD sum rules \cite{braunhalperin,aliev,ballbraun,kou,khodj1,khodj2,zwicky} 
have been applied to obtain the theoretical predictions for such processes.  
The $B$-decays into two currents are related to correlation functions containing $B$-meson in the initial state 
and one can calculate them in QCD via the $B$-meson distribution amplitudes \cite{grozin,neubertgrozin,japan,braunkorchemsky}. 

\begin{figure}[b!]
\begin{tabular}{cc}
\includegraphics[width=5cm]{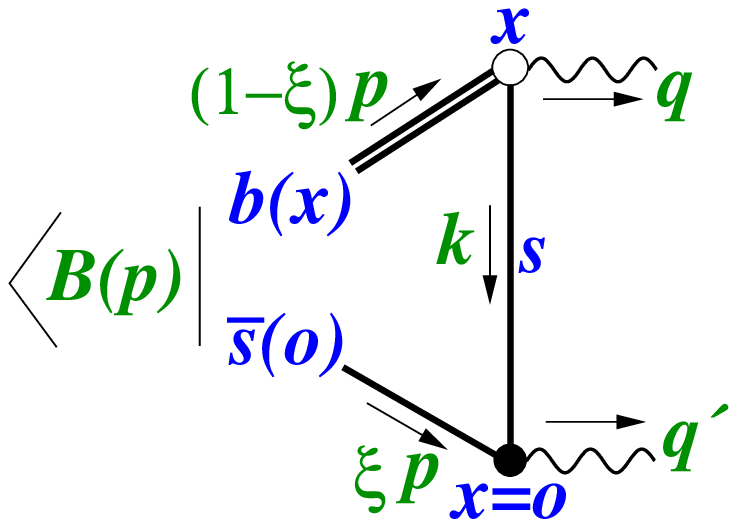}\qquad  &  \qquad \includegraphics[width=5cm]{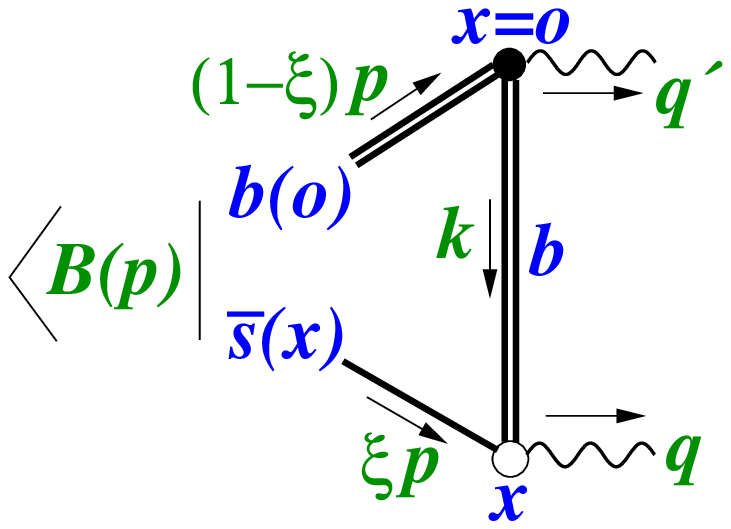}\\
(a) & (b)\\
\end{tabular}
\caption{\label{Fig:1} 
Feynman diagrams describing: 
(a) $\langle 0| T\{\bar s(x)b(x), \bar s(0)s(0)\}|B(p\rangle)$, (b) $\langle 0| T\{\bar s(x)b(x), \bar b(0)b(0)\}|B(p)\rangle$.}
\end{figure}

The form factors of our interest are complicated objects defined by the following amplitude 
\begin{eqnarray}
\label{def}
T(q,q'|p)=i\int dx\,  e^{i q x} \langle 0| T\{\bar s(x){\cal O}b(x), j(0)\}|B(p)\rangle = F(q^2,q'^2), \qquad p=q+q'.
\end{eqnarray}
In practically interesting cases, the weak current $\bar s(x){\cal O} b(x)$ contains some Dirac matrix ${\cal O}$, and 
$j(y)$ may be the electromagnetic current 
\begin{eqnarray}
\label{jem}
j_\mu^{\rm e.m.}(y)=Q_b \bar b(y)\gamma_\mu b(y) + Q_s \bar s(y) \gamma_\mu s(y).
\end{eqnarray}
Respectively, the amplitude (\ref{def}) involves a number of Lorentz structures and form factors $F_i(q^2,q'^2)$; each of these form factors receives  
contributions from two diagrams shown in Fig.~\ref{Fig:1}. The diagram Fig.~\ref{Fig:1}(a) gives the dominant contribution 
in the heavy-quark limit \cite{korchemsky,mn} and we shall therefore discuss this diagram. 

For the sake of argument, we omit here the Lorentz and the spinor indices and consider field theory with scalar partciles. 
Such an approach allows us to avoid techncal complications and to concentrate on the conceptual aspects \cite{lucha1,lucha2,km2018}. 
Still, we keep the QCD notations of the quark fields $b$ and $s$ and denote $\bar b$ and $\bar s$ the complex conjugate fields.
Here $b$ denotes the $b$-quark field and $s$ denotes the light-quark field ($d$, $u$ or $s$). For scalar particles, the form factor 
corresponding to Fig.~\ref{Fig:1}(a) may be written as 
\begin{eqnarray}
\label{ff}
F(q^2,q'^2) = i \int dx\,dk\, e^{i(q-k) x} D_s(k)\langle 0|\bar s(0)b(x)|B(p)\rangle,
\end{eqnarray}
where $D_s(k)$ is the propagator of the scalar light $s$``quark''.

\section{Expansion of the $B$-meson wave function near the light cone}
We shall use the following general representation for a non-local 
two-quark distribution in the $B$-meson (the Bethe-Salpeter amplitude) \cite{braunhalperin,km2018}
\begin{eqnarray}
\label{BS}
\langle 0|\bar s(0)b(x)|B(p)\rangle=\int\limits_0^1 d\xi e^{-i p x (1-\xi)}\left[\phi_B(\xi)+x^2\phi_{B(2)}(\xi)+\dots\right]
\end{eqnarray}
The $\xi$-integration runs from $0 < \xi < 1$ as follows from the general properties of Feynman diagrams in field theory. 
Here $\phi_B(\xi)$ is the light-cone (LC) distribution amplitude (DA) of the $B$-meson; the functions $\phi_{B(n)}(\xi)$ denote the distribution 
amplitudes of higher twist. The LC DA corresponding to $x^2=0$ term in Eq.~(\ref{BS}) is normalized as  
\begin{eqnarray}
\label{fB}
\langle 0|\bar s(0)b(0)|B(p)\rangle=\int\limits_0^1 d\xi \phi_B(\xi)=f_B,
\end{eqnarray}
with $f_B$ the leptonic constant of the $B$-meson. 
The DA $\phi_B(\xi)$ as well as the higher-order DAs have the following properties: \\
(i) they are peaked in the region $\xi\sim \Lambda_{\rm QCD}/m_b$, and \\
(ii) in field theory with scalar particles, they vanish at the end-points $\phi_B(\xi=0)=\phi_B(\xi=1)=0$. 

Because of the peaking of the $B$-meson DAs at small $\xi$, in many applications one can run the $\xi$-integration up to $\infty$; 
such integration limits emerge in the DAs of heavy-quark effective theory \cite{grozin,neubertgrozin}. 

Physically meaningful and theoretically consistent parametrizations of the $B$-meson DAs may be taken in the forms 
\begin{eqnarray}
\label{phi1}
&&\phi_B(\xi)\sim \xi(1-\xi)^n,\\
\label{phi2}
&&\phi_B(\xi)\sim \exp\left\{-\frac{1}{\beta_B\, m_b}\left(\frac{m_b^2}{1-\xi}+\frac{m^2}{\xi}-m_b^2\right)\right\}, 
\quad \beta_B\sim \Lambda_{\rm QCD}
\end{eqnarray}
Particularly convenient for our purpose is the first parametrization; it allows us to obtain explicit analytic representation 
for the form factor and to study its analytic properties. For $B$-meson, $n$ should be large, of order 
$m_b/\Lambda_{\rm QCD}$.


\section{Form factor: the light-cone expansion}
The $x^2$-expansion of the BS amplitude (\ref{BS}) generates the corresponding expansion of 
the form factor (\ref{ff1}): The term proportinal to $\phi_B$ in (\ref{BS}) corresponds to $x^2=0$ and thus describes 
the light-cone configuration of the quarks inside the $B$-meson. Hereafter we write $\phi(\xi)$ instead of $\phi_B(\xi)$. 
Its contribution to the form factor (\ref{def}) is easy to calculate:  
\begin{eqnarray}
\label{fflc}
F(q^2,q'^2)=
\frac{1}{(2\pi)^4}\int\limits_0^1 d\xi \phi(\xi) \int dx e^{iqx} e^{-i(1-\xi)px}\int dk e^{-ikx}
\frac{1}{m_s^2-k^2-i0}=
\int\limits_0^1 \frac{d\xi \,\phi(\xi)}{m_s^2-(q'-\xi p)^2}
\end{eqnarray}
Taking into account that $(p-q')^2=q^2$, and thus $2q'p=p^2+q'^2-q^2$, we obtain 
\begin{eqnarray}
\label{k2}
-k^2=\xi(1-\xi)M_B^2-q^2\xi-q'^2(1-\xi)
\end{eqnarray}
and, finally, the form factor takes the form 
\begin{eqnarray}
\label{ff1}
F(q^2,q'^2) = \int\limits_0^1 d\xi \phi(\xi) \frac{1}{m^2+M_B^2\xi(1-\xi)- q^2\xi - q'^2(1-\xi) - i 0}
\end{eqnarray}
We have used the Feynman quark propagator for the calculation of the hadron form factor. This procedure 
is consistent if the bulk of the form factor comes from the integration region where quark is far from it mass shell. 

Taking into account the peaking of $\phi(\xi)$ near $\xi\sim \Lambda_{\rm QCD}/m_b$, we find that 
the propagating light quark is highly virtual, $k^2\sim -\Lambda_{\rm QCD}m_b$, in the region of the external momenta satisfying 
$q^2 < M_B^2$ and $q'^2 < \Lambda_{\rm QCD}m_b$. In this region, the propagator is positive and large in the $\xi$-integration region and 
our calculation of the $B$-meson form factor is trustable.\footnote{Eq.~(\ref{ff1}) determines the form factor as an analytic function for all values of $q^2$ and $q'^2$, but 
QCD confinement limits the region of the external momenta, where the form factor obtained from Eq.~(\ref{ff1}) provides
realistic predictions for the $B$-meson form factor.} 
Another region of the external momenta where Eq.~(\ref{ff1}) gives trustable predictions for the $B$-meson form factor corresponds 
to those $q^2$ and $q'^2$ where the denominator in (\ref{ff1}) is negative and large in the full $\xi$-integration region and thus 
the light $s$-quark is far from its mass shell.

Let us turn to those contributions to the form factor that are generated by higher twist effects 
(i.e. higher powers of $x^2$) in the expansion (\ref{BS}). To calculate such contributions, 
it is convenient to substitute in Eq.~(\ref{fflc}) $x_\alpha=-i\frac{\partial}{\partial k_\alpha} e^{ikx}$. 
By performing the parts integration, the $k_\alpha$-derivative acts on the $s$-quark propagator. Therefore in the region where the 
quark propagator is highly virtual, $k^2\sim -\Lambda_{\rm QCD}m_b$, the contributions of higher-twist terms are power-suppressed 
$(x^2)^n\to (\Lambda_{\rm QCD}/m_b)^n$ compared to leading-twist LC term. 

The analytic properties of the form factor (\ref{ff}) depend on the singularities of the denominator in the integral (\ref{ff}). 
Therefore, the structure of singularities is the same in the LC and in the higher-twist contributions, 
and we will concentrate on the analytic properties of the LC form factor (\ref{ff1}).

Before studing these properties, however, we discuss some model-independent predictions for the behaviour 
of the form factor in specific regions of $q^2$ and $q'^2$. 

\section{\label{Sect:4}Model-independent predictions for the form factor}
In some regions of the momentum transfers $q^2$ and $q'^2$, model-independent features 
of the form factor $F(q^2,q'^2)$ may be established. 
To highlight these properties, one needs to make use of only one essential property of the LC wave function 
$\phi(\xi)$: namely, its peaking 
in the end-point region $\xi\simeq \Lambda_{\rm QCD}/m_b$. This peaking reflects the fact that the major part of the heavy-meson momentum 
is carried by the heavy quark. 

We emphasize below three cases which may have impact on constraining the phenomenological models for the form factors of this kind. 
\begin{itemize}
\item[a.] $q'^2=q^2$, and $|q^2|\gg M_B\Lambda_{\rm QCD}$. In this case, we can neglect the term $m^2+M_B^2\xi(1-\xi)$ in the denominator
and by virtue of (\ref{fB}) obtain 
\begin{eqnarray}
F(q^2,q^2)\simeq \frac{f_B}{-q^2}. 
\end{eqnarray}
\item[b.] $q'^2\simeq \Lambda_{\rm QCD}^2$. Then, 
\begin{eqnarray}
\label{pole}
F(q^2,q'^2\simeq \Lambda_{\rm QCD}^2) \simeq \int\limits_0^1 d\xi \phi(\xi) \frac{1}{\xi(M_B^2(1-\xi)-q^2)}\simeq
\frac{1}{(M_B^2-q^2)}\int\limits_0^1 d\xi \frac{\phi(\xi)}{\xi}.
\end{eqnarray}
Therefore, for $q'^2\simeq \Lambda_{\rm QCD}^2$, the $q^2$-dependence of the form factor in a broad 
range of $q^2$ has a monopole behavior with the pole location at $q^2\simeq M_B^2$. 
Corrections to the precise pole location are of order $\Lambda_{\rm QCD}m_b$. 
This property is physically transparent and corresponds to the contribution to the amplitude coming from the beauty meson 
with the appropriate quantum numbers. It therefore gives a theoretical justification of the monopole model for the 
form factor in a broad range of positive $q^2$ and small positive $q'^2$. 
\item[c.]
In the region $M_B\Lambda_{\rm QCD} \ll q'^2\simeq M_B^2$ and for all $0 < q^2 < M_B^2$, 
the form factor behaves like a monopole function in $q'^2$: 
\begin{eqnarray}
\label{pole2}
F(0<q^2< M_B^2, q'^2\simeq M_B^2) \simeq \int\limits_0^1 d\xi \phi(\xi) \frac{1}{-q'^2(1- \xi)}=
\frac{f_B}{-q'^2}\left(1+O(\Lambda_{\rm QCD}/m_b)\right).
\end{eqnarray}
\end{itemize}


\begin{figure}[b!]
\begin{tabular}{c}
\includegraphics[width=9cm]{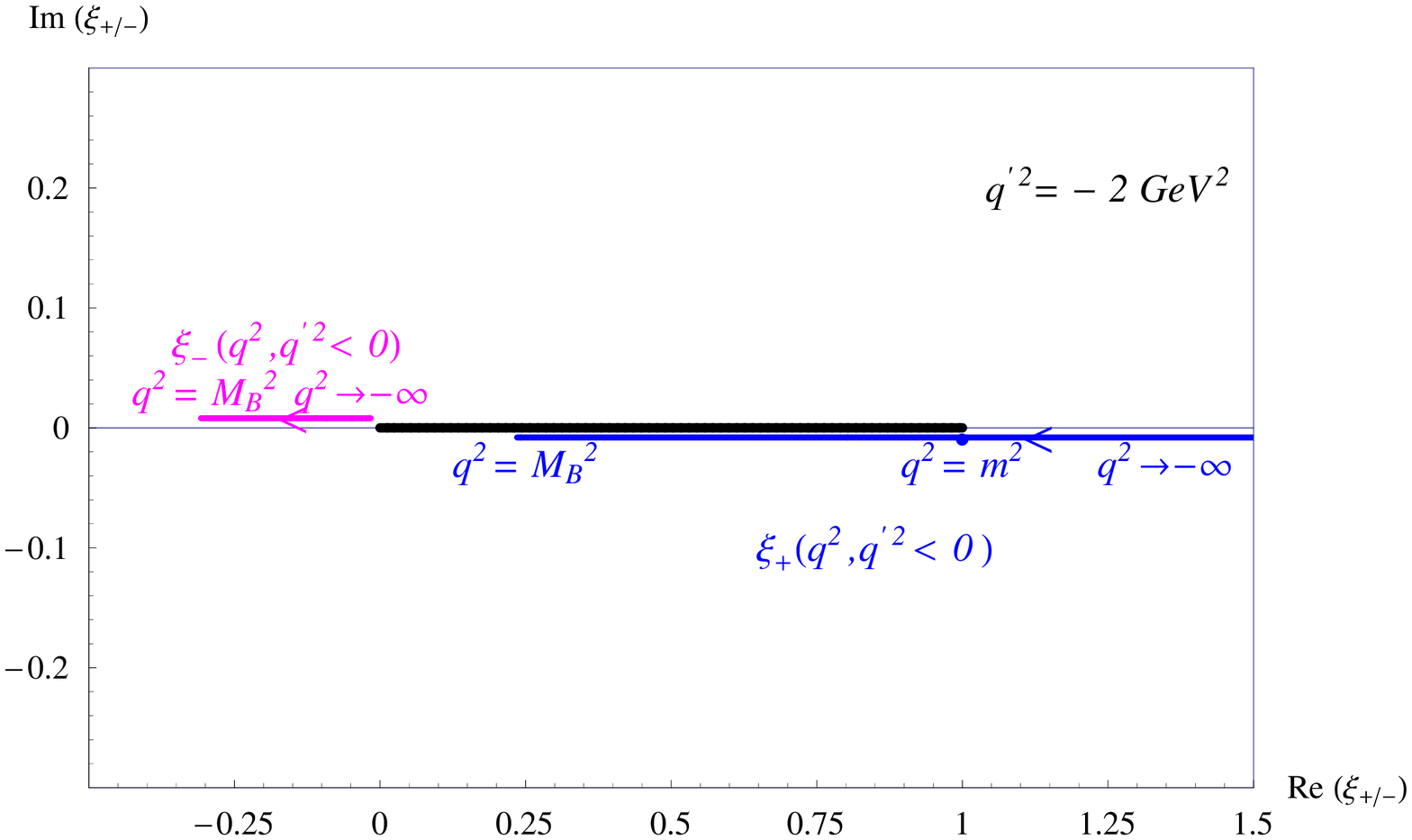}\\
(a)\\
\includegraphics[width=9cm]{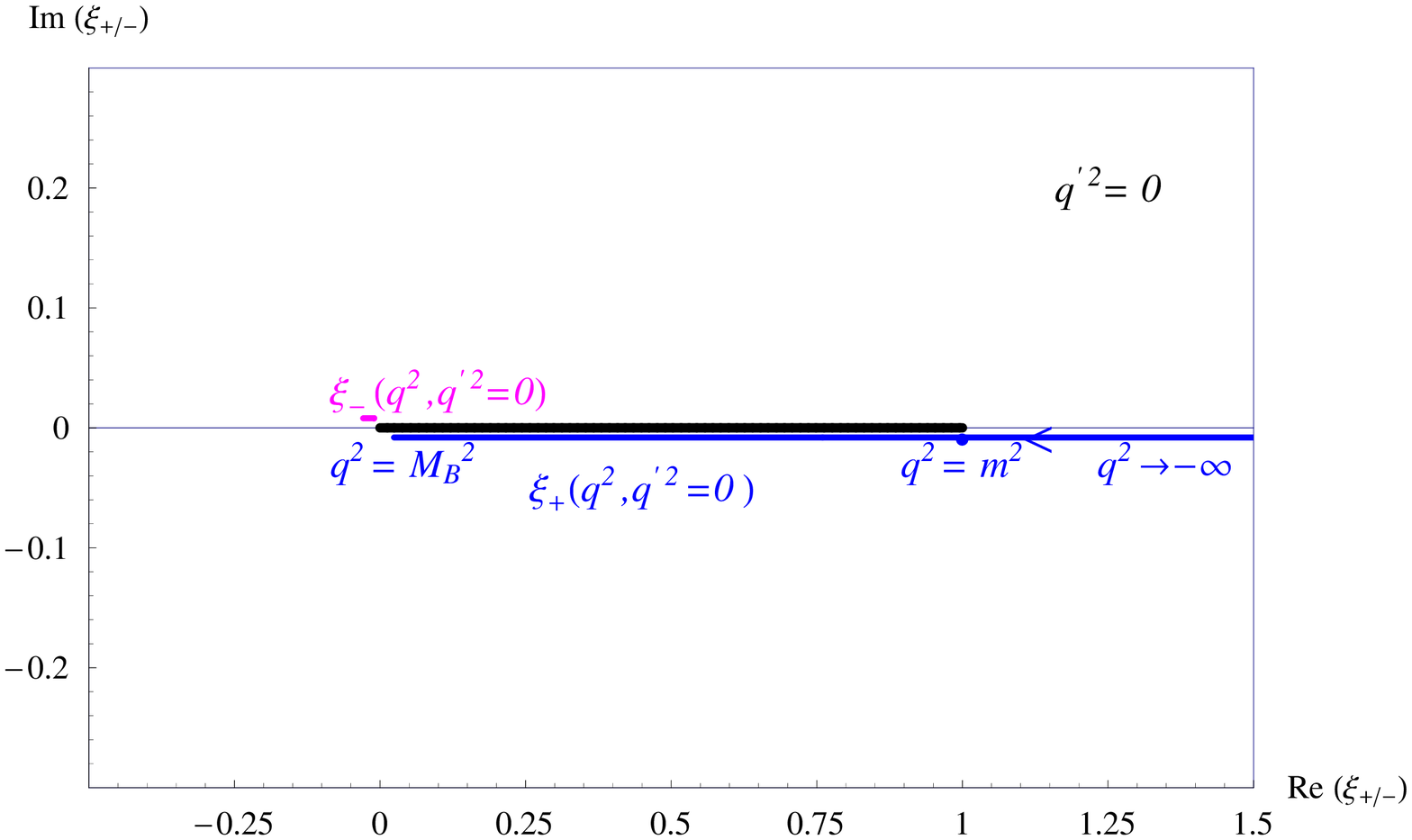} \\
(b)\\
\includegraphics[width=9cm]{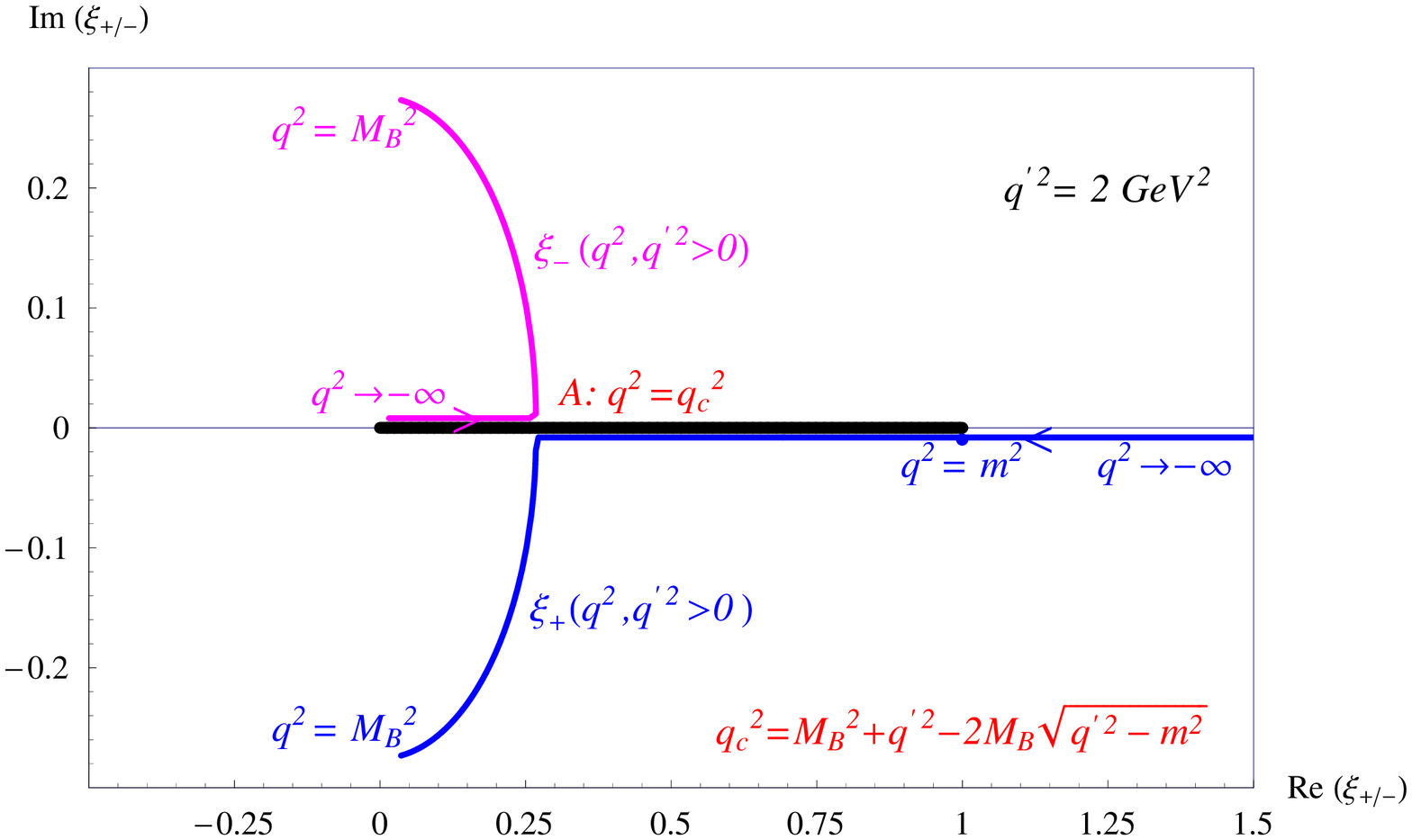}\\
(c)\\
\end{tabular}
\caption{\label{Plot:1} 
Trajectories of the singularities of the integrand, $\xi_+(q^2,q'^2)$ and $\xi_-(q^2,q'^2)$, vs $q^2$ at a fixed value $q'^2$  
in the complex $\xi$-plane. 
(a): $q'^2<0$ [$q'^2=-2$ GeV$^2$]. 
(b): $q'^2=0$. 
(c): $q'^2>m^2$ [$q'^2=2$ GeV$^2$]; here $A$ denotes the critical value of $q^2$, at which $\xi_+$ and $\xi_-$ start to move away from the real axis. 
This happens for $q^2>q_c^2$, where $q_{c}^2=M_B^2+q'^2-2 M_B \sqrt{q'^2-m^2}$. For $q^2>q_{c}^2$, $\xi_+$ and $\xi_-$ are complex conjugate 
numbers, $\xi_+=\xi_-^*$. Since $q_{c}^2< (M_B-\sqrt{q'^2})^2$, $q_{c}^2$ lies beyond the physical region of the $B$-decay.}
\end{figure}

\section{Form factor: The analytic properties}
We now turn to the analysis of the analytic properties of the form factor (\ref{ff1}).
Because of the Feynman propagator in the integrand, this $\xi$-integral is a contour integral in the complex $\xi$-plane along the cut 
located at $0 < \xi < 1$. Consequently, the analytic properties of the form factor are determined by the motion of the 
singularities of the denominator
in the complex $\xi$-plane depending on $q^2$ and $q'^2$, and the analytic properties of the form factor in the theory with 
scalar ``quarks'' coincide with the analytic properties of the form factor in QCD.

Clearly, if $q^2<m^2$ and $q'^2<m^2$, the denominator is positive in the $\xi$-integration region; the $-i 0$ addition may be safely 
omitted, and the form factor is the real function in this kinematic domain. However, if one of the variables, $q^2$ or $q'^2$, lies 
in the timelike region above $m^2$, the form factor acquires the imaginary part. As we shall see, $F(q^2,q'^2)$ 
has logarithmic branch points 
in this kinematical region and the analysis of its analytic properties becomes a more complicated problem. 
To get to the right branch of the logarithmic cut, it is convenient to keep a small imaginary addition in $m$, 
the mass of the propagating particle, i.e. to replace $m^2\to m^2-i 0$. 

The denominator is a quadratic function of the variable $\xi$ and may be written as follows
\begin{eqnarray}
\label{den}
m^2+M_B^2\xi(1-\xi)-q^2\xi-q'^2(1-\xi)= - M_B^2 \left[\xi-\xi_+(q^2,q'^2)\right]\left[\xi-\xi_-(q^2,q'^2)\right],
\end{eqnarray}
where 
\begin{eqnarray}
\label{xipm}
\xi_{\pm}(q^2,q'^2)=\frac{M_B^2+q'^2-q^2\pm\sqrt{4 M_B^2m^2+(M_B^2-q'^2+q^2)^2-4M_B^2 q^2}}{2M_B^2}.
\end{eqnarray}
The form factor Eq.~(\ref{ff1}) may be rewritten as (for brevity, we omit the arguments $q^2$ and $q'^2$ in $\xi_\pm$):
\begin{eqnarray}
\label{F}
F(q^2,q'^2)=-\frac{1}{M_B^2}\frac{f(\xi_+)-f(\xi_-)}{\xi_+-\xi_ {-}}
\end{eqnarray}
with 
\begin{eqnarray}
\label{f1}
f(\xi_0)=\int\limits_0^1\frac{d\xi \phi(\xi)}{\xi-\xi_0}.
\end{eqnarray}
Here $f(\xi_0)$ is the analytic function of the complex variable $\xi_0$ with the cut along the real axis from 0 to 1. 
Recall, that both $\xi_+$ and $\xi_-$ have imaginary parts fully determined by the small negative imaginary addition in the quark mass $m$; 
this imaginary addition fully determines the location of the complex variable $\xi_\pm$ respective to the cut in the function $f$.  
The analytic properties of the form factor are then fully determined by the trajectories of the 
logarithmic branch point $\xi_+$ and $\xi_-$ 
vs $q^2$ and $q'^2$. These trajectories are shown in Fig.\ref{Plot:1}. 

In those region of the momentum transfers, for which both $\xi_+$ and $\xi_-$ lie away from the range from 0 to 1 on the real axis, 
the calculation is straightforward. 

However, if the values of $\xi_+$ or $\xi_-$ belong to the interval from 0 to 1 on real axis (in the limit $\epsilon\to 0$), 
it is convenient to isolate the singular point in the integrand of (\ref{f1}): namely, for $0 < {\rm Re}\, \xi_0 <1$ and $|{\rm Im}\, \xi_0|\to 0$, 
we use the following representation for $f(\xi_0)$:   
\begin{eqnarray}
\label{f}
f(\xi_0)&=&\int\limits_0^1 d\xi \frac{\phi(\xi)-\phi(\xi_0)}{\xi-\xi_0}+\phi(\xi_0)\int\limits_0^1 \frac{d\xi}{\xi-\xi_0}
\nonumber\\&=&\int\limits_0^1 d\xi \frac{\phi(\xi)-\phi(\xi_0)}{\xi-\xi_0}+\phi(\xi_0)\log\left(\frac{\xi_0-1}{\xi_0}\right).
\end{eqnarray}
The first integral here is a regular function without a singularity in the integration region, whereas the second logarithmic term fully determines 
the logarithmic singular part of the form factor with the cut on the real axis at $0 < \xi_0 < 1$.
Let us emphasize that the form factor $F(q^2,q'^2)$ of Eq.~(\ref{F}) 
has a nonzero imaginary part if at least one of $\xi_+(q^2,q'^2)$ or $\xi_-(q^2,q'^2)$ lies on the boundary of this cut. 

\newpage
\section{Numerical examples}
In this Section, we address two problems: First, we illustrate the general structure of the form factor by the calculation 
in an explicit model for the LC DA (\ref{phi1}) with $n=9$; Second, we demonstrate the sensitivity of the form factor to the specific shape 
of the LC DA, comparing the results obtained with two different LC DAs. For numerical estimates we need input parameters such 
as ``quark'' masses and the DA of ``$B$-meson'' in our model with scalar constituents. We will make use of the realistic 
values of quark and meson masses $m_s=0.1$ GeV, $m_b=4.2$ GeV, and $M_B=5.27$ GeV. Throughout this Section, 
we normalize the LC DA as $\int_0^1\phi(\xi)d\xi=1$. 

\subsection{Form factor in a broad range of momentum transfers}
\begin{figure}[b!]
\begin{tabular}{cc}
\includegraphics[width=8cm]{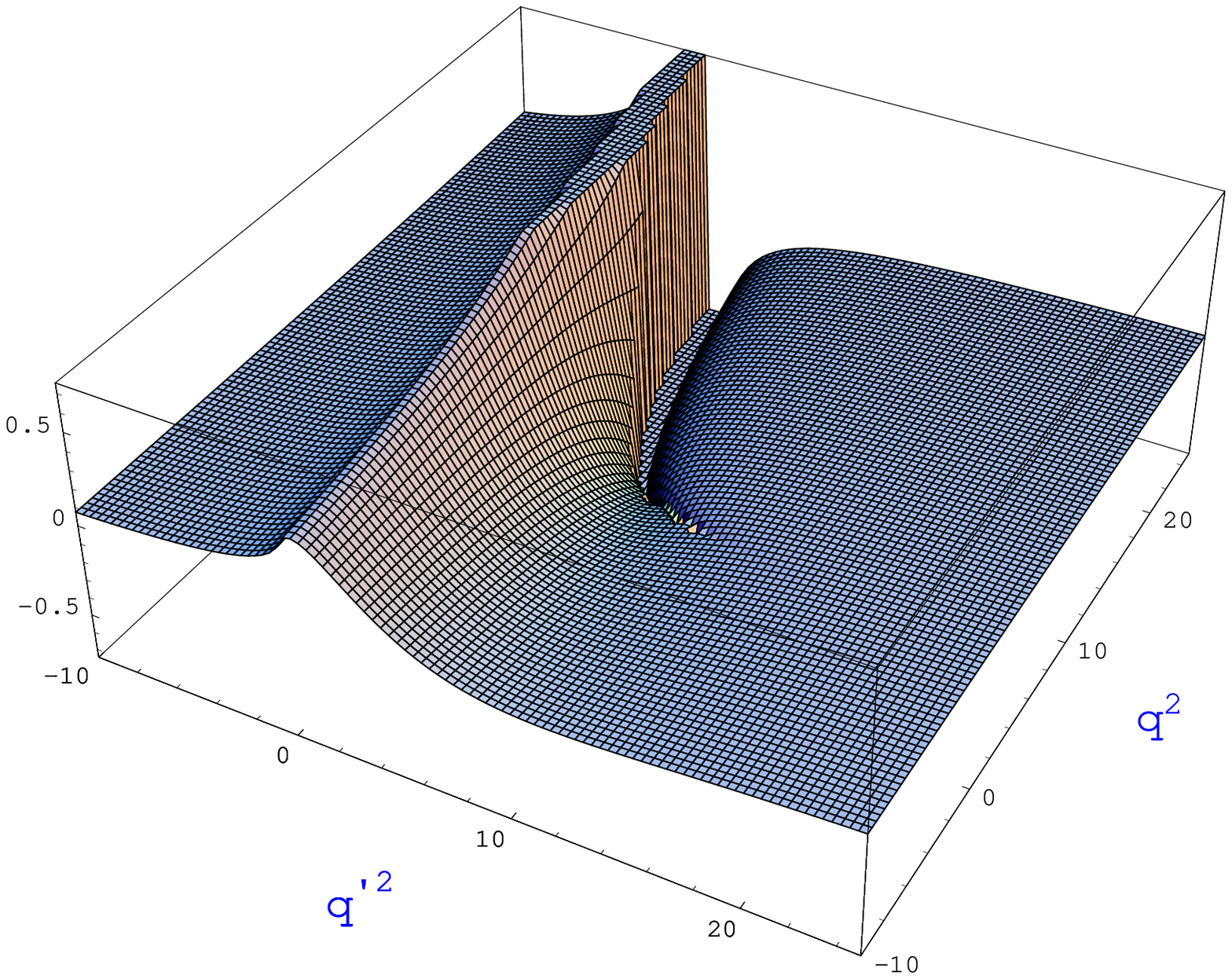} & \includegraphics[width=8cm]{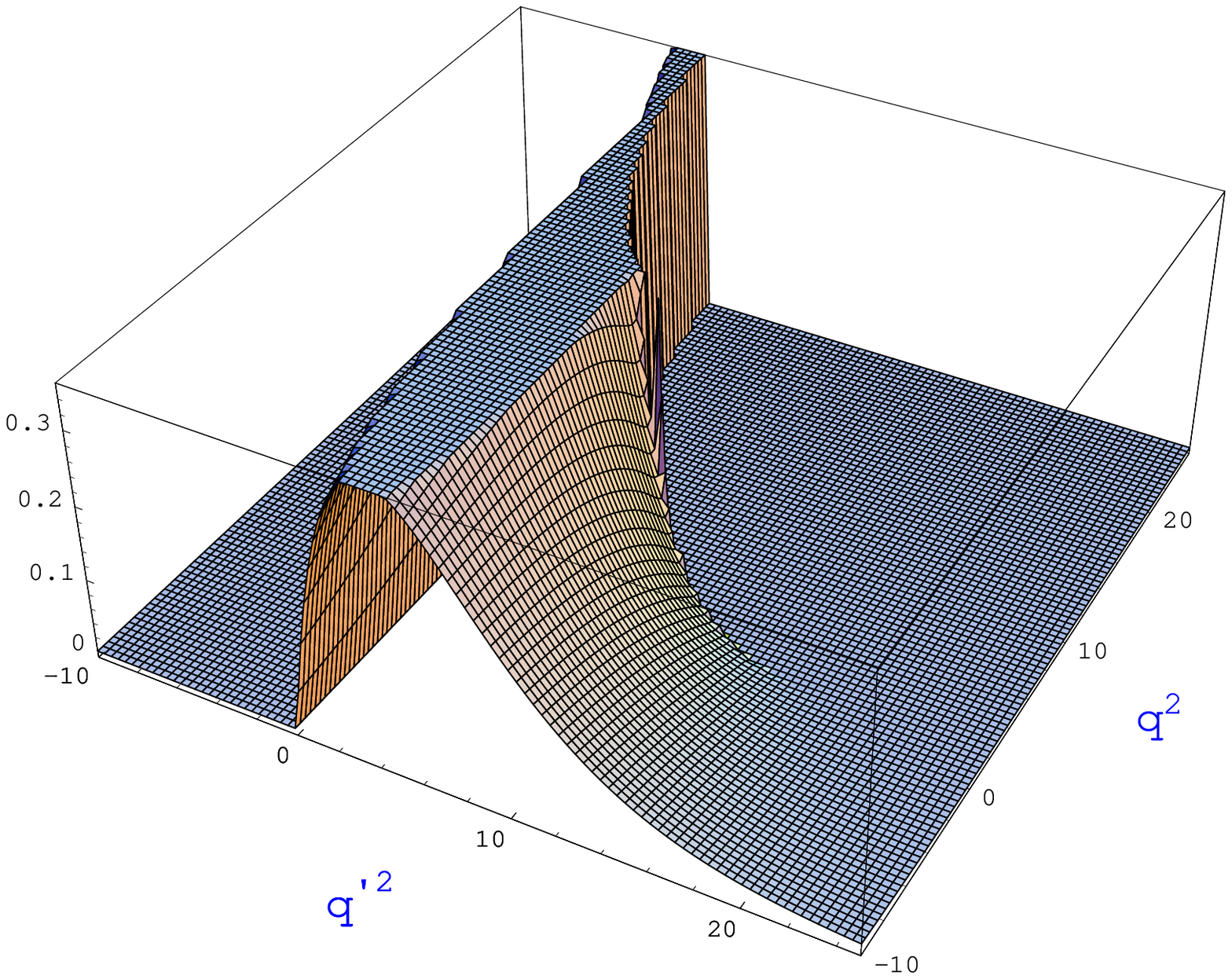}\\
(a) & (b) \\
\end{tabular}
\caption{\label{Plot:2}  
$F(q^2,q'^2)$ in a broad range of two variables $-10 \mbox{ GeV}^2 < q^2,q'^2 < M_B^2$: 
the real (a) and the imaginary (b) parts. 
The physical region of the $B$-decay is narrower and is determined by the 
following conditions: $4m_l^2 < q^2$, $4m_l^2 <q'^2$, and $\sqrt{q^2}+\sqrt{q'^2}\le M_B$.}
\end{figure}

\begin{figure}[b!]
\begin{tabular}{cc}
\includegraphics[width=8cm]{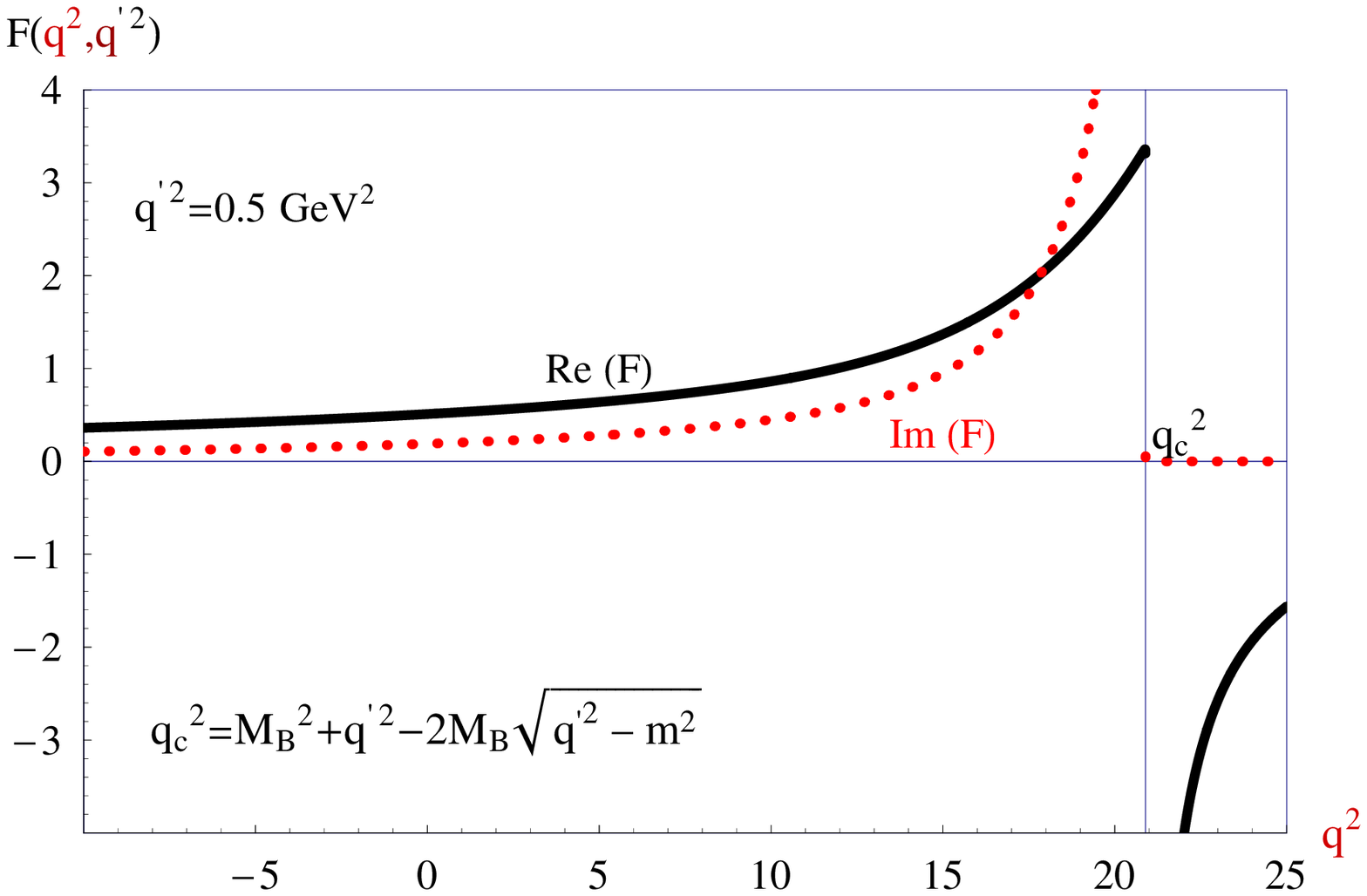} & \includegraphics[width=8cm]{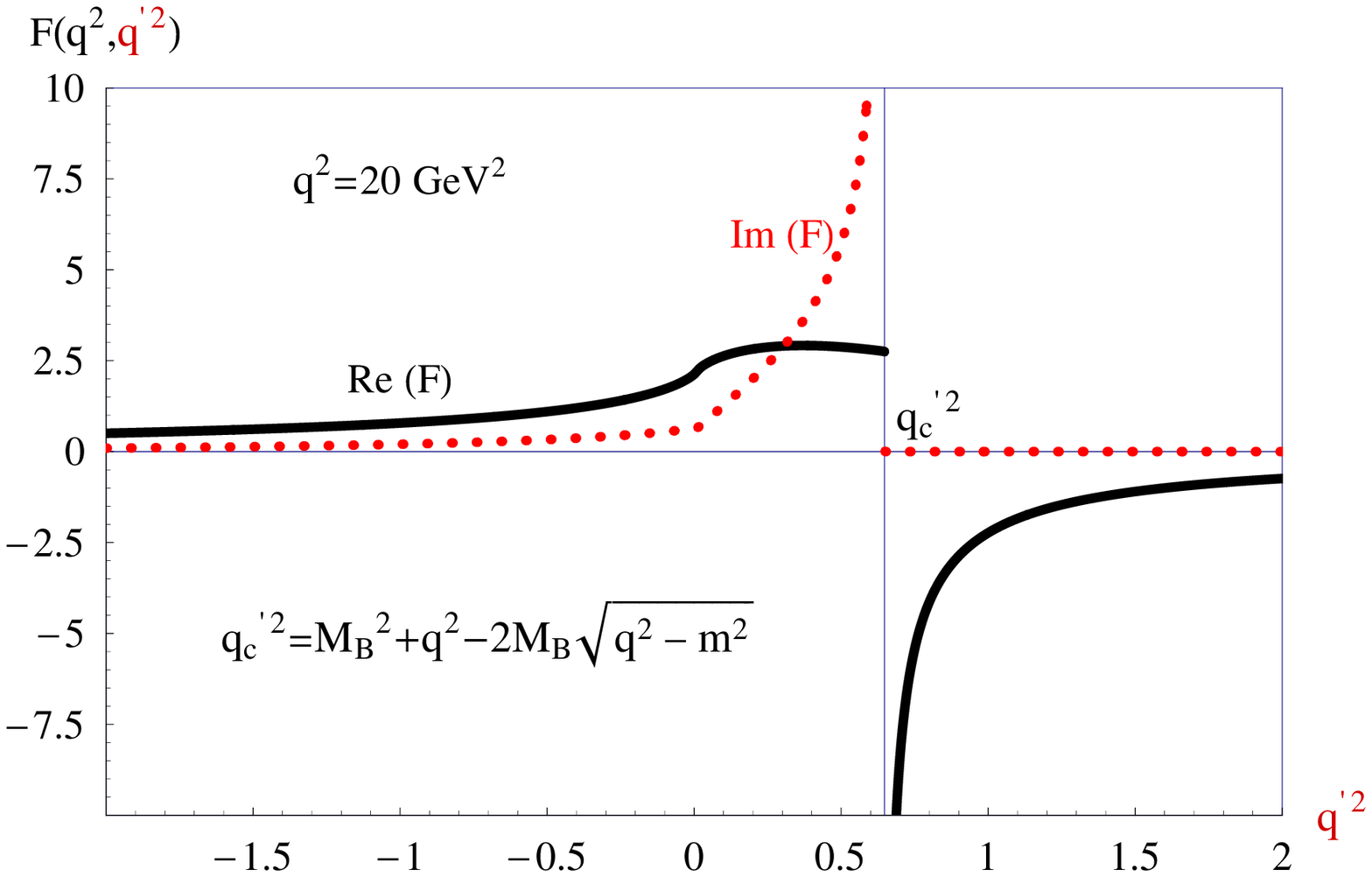}\\
(a) & (b)\\
\end{tabular}
\caption{\label{Plot:3}  
$F(q^2,q'^2)$ as function of one of the momentum squared at a fixed value of the other momentum squared: 
(a) $F(q^2,q'^2=0.5\mbox{ GeV}^2)$, (b) $F(q^2=20\mbox{ GeV}^2,q'^2)$. }
\end{figure}

Figure \ref{Plot:2} shows the form factor as function of two variables $q^2$ and $q'^2$ 
in a broad range of momentum transfers for the $B$-meson LC DA (\ref{phi1}) with $n=9$; the form factor in this case may 
be obtained as an explicit function containing polynomials and logarithms. 
The physical $B$-decay region is restricted to timelike $q$ and $q'$ such that $\sqrt{q^2}+\sqrt{q'^2}\le M_B$; it 
corresponds to a part of the region displayed in the plot. 

Figure \ref{Plot:3} shows $F(q^2,q'^2)$ as function of one momentum squared at a fixed value of the other momentum squared. 
Recall that the Feynman rules prescribe a small imaginary addition $-i\epsilon$ in the propagator, so, more rigorously, we 
have the function of three variables, $F(q^2,q'^2|\epsilon)$. The form factor is then defined as 
$F(q^2,q'^2)\equiv F(q^2,q'^2|\epsilon\to 0)$. 

The limit $\epsilon\to 0$ may be taken without problem for all momentum transfers, except for the family of 
``critical'' points $q^2$ and $q'^2$ for which both roots $\xi_+$ and $\xi_-$ pinch 
in the limit $\epsilon\to 0$, i.e., those $q^2$ and $q'^2$ which satisfy the condition $4 M_B^2m^2+(M_B^2-q'^2+q^2)^2-4M_B^2 q^2=0$. 
This condition determines a ``critical'' curve (a parabola) in the $q^2$-$q'^2$-plane, where the function $F(q^2,q'^2|\epsilon)$ 
has an essentail singularity 
and the limiting values $\lim\limits_{\epsilon\to 0}F(q_c^2,q'^2|\epsilon)$ and $\lim\limits_{q^2\to q_c^2} F(q^2,q'^2|\epsilon=0)$ 
differ from each other. (If we fix $q^2$ and consider the form factor as the function of $q'^2$, then
$\lim\limits_{\epsilon\to 0}F(q^2,q_c'^2|\epsilon)$ and $\lim\limits_{q'^2\to q_c'^2} F(q^2,q'^2|\epsilon=0)$ 
differ from each other).

Fortunatley, this critical curve lies beyond the physical decay region and therefore the precise way of taking 
the limit is not crucial. If one chooses the procedure of taking first $\epsilon\to $, then one has the picture shown in Fig.~\ref{Plot:3}: 
The function $F(q^2,q'^2)$ is discontinuous on the ``critical'' line and has different left and right limiting values. 
Moreover, the imaginary (the real) part of the form factor 
diverges on the left (on the right) from the critical value of the momentum squared. For instance, in Fig.~\ref{Plot:3}(a) 
${\rm Im}\,F(q^2,q'^2)\propto (q^2-q_c^2)^{-1/2}$ for $q^2<q_c^2$, and ${\rm Re}\,F(q^2,q_0'^2)\propto (q^2-q_c^2)^{-1/2}$ for $q^2>q_c^2$, where 
$q_c^2=M_B^2+q'^2-2MB\sqrt{q'^2-m^2}$. 
This behaviour is easy to understand from the property of the function $f(\xi_0)$ in Eq.~(\ref{f}): namely, 
$f(\xi_0)$ has a cut along the real axis from 0 to 1, and $\xi_-$ lies on the upper boundary of the cut, 
whereas $\xi_+$ is on the lower boundary. 

Let us consider the limit $q^2\to q_c^2$ from below. From Eq.~(\ref{xipm}) it follows that 
${\rm Re}\,(\xi_+)\to {\rm Re}\,(\xi_-)\propto \sqrt{q^2-q_c^2}$. 
The real part, ${\rm Re}\, f(\xi)$, is continuous on the real axis, such that ${\rm Re}\, f(\xi_+)-{\rm Re}\, f(\xi_-)$ 
vanishes as ${\rm Re}\,(\xi_+)\to {\rm Re}\,(\xi_-)$, 
whereas the imaginary part, ${\rm Im}\,f(\xi_+)$ has a finite discontinuity, such that 
${\rm Im}\, f(\xi_+)-{\rm Im}\, f(\xi_-)$ remains finite as $\xi_+\to \xi_i$. As the result, 
${\rm Im}\,F(q^2,q_0'^2)\sim (q^2-q_c^2)^{-1/2}$ whereas ${\rm Re}\, F(q^2,q_c'^2)$ has a finite limiting value.

If we consider the limit $q^2\to q_c^2$ from above, the situation changes: now ${\rm Im}\,(\xi_+)-{\rm Im}\,(\xi_-)\propto \sqrt{q^2-q_c^2}$, 
leading to ${\rm Re}\, F(q^2,q'^2)\propto -(q^2-q_c^2)^{-1/2}$, whereas ${\rm Im}\, F(q^2,q'^2)\equiv 0$ for $q^2>q_c^2$.

A very similar picture is seen in the $q'^2$-dependence of the form factor $F(q^2,q'^2)$ at a fixed $q^2$ in Fig.~\ref{Plot:3}(b).

\subsection{Sensitivity of the form factor to the DA shape}

\begin{figure}[b!]
\begin{tabular}{cc}
\includegraphics[width=7cm]{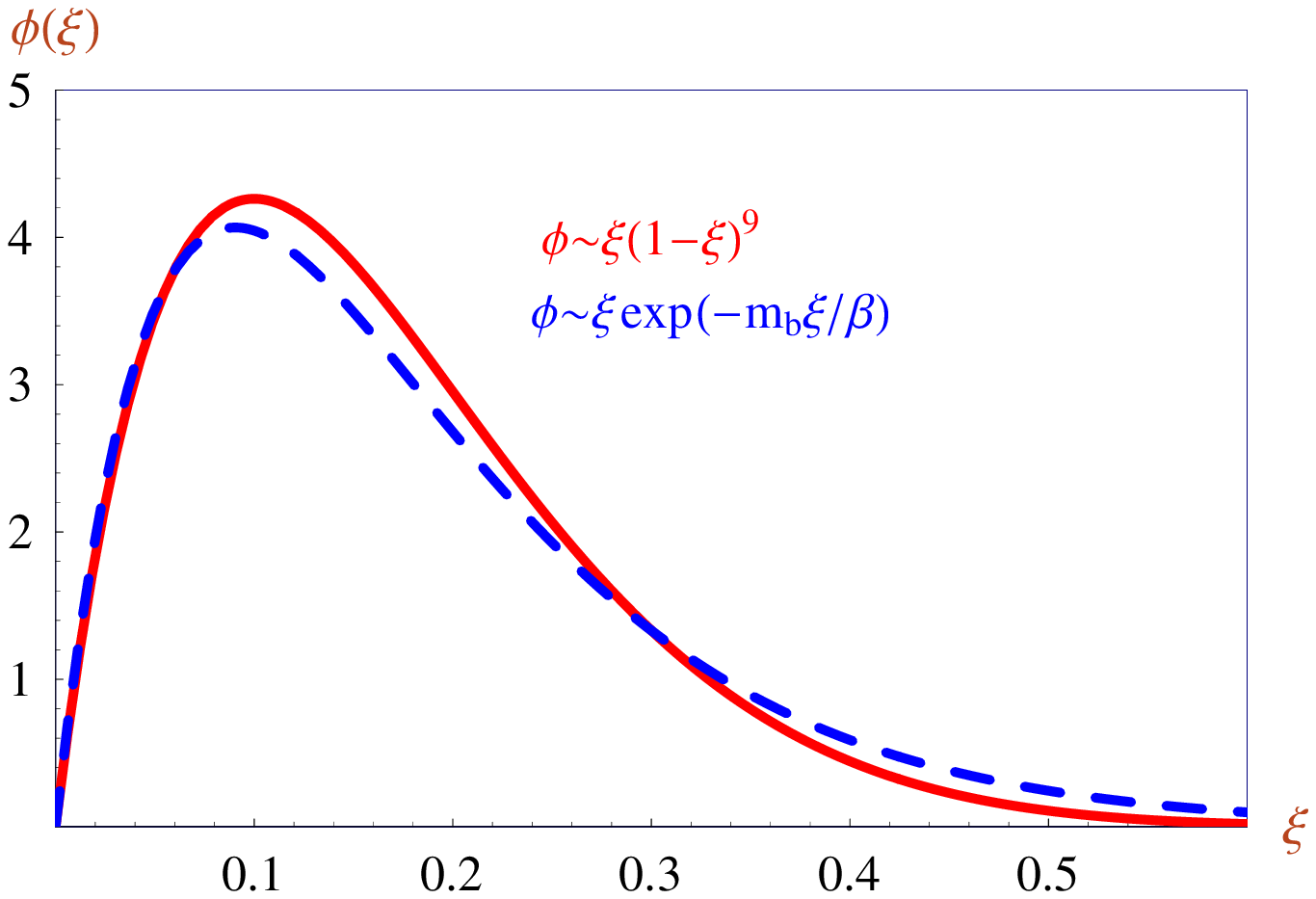} & \includegraphics[width=7cm]{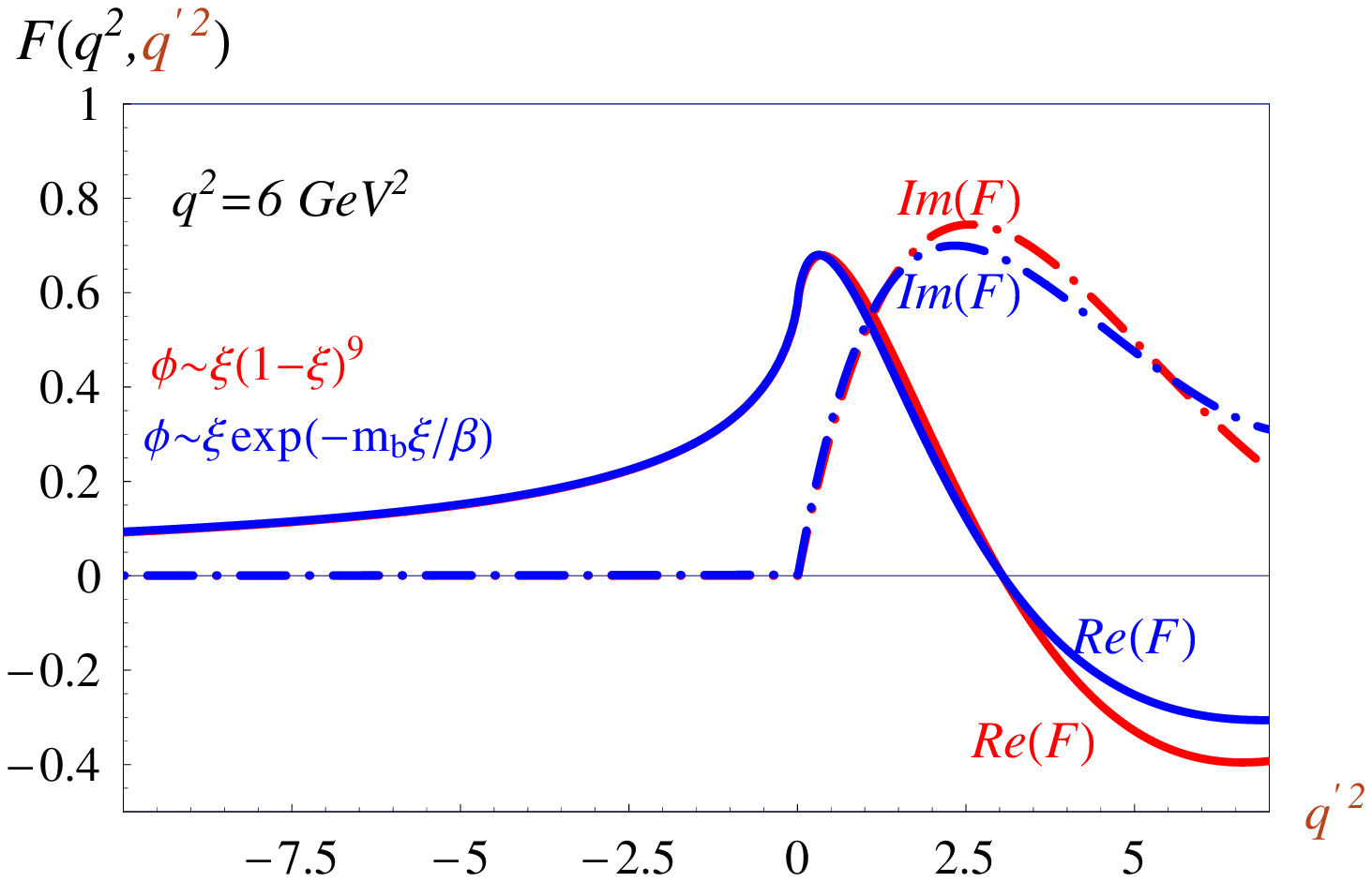}\\
(a) & (b)\\
\end{tabular}
\caption{\label{Plot:4}  
(a) Two different parametrizations of the $B$-meson DA $\phi(\xi)$: Eq.~(\ref{DA1}) with $n=9$ (red) and Eq.~(\ref{DA2}) 
and (b) the corresponding $F(q^2,q'^2)$ vs $q'^2$ for a fixed $q^2=6$ GeV$^2$. }
\end{figure}
As seen from the results of Section~\ref{Sect:4}, in some regions of the momentum transfers the form factor 
is expected to be sensitive to the 
DA shape, in particular, to the value of its first inverse moment 
$\langle{\xi^{-1}}\rangle= \int\limits_0^1 \frac{\phi(\xi)}{\xi} d\xi$. 
We study the sensitivity of the form factor to the specific shape of the DA for three cases  
\begin{eqnarray}
\label{DA1}
&&\phi_1(\xi)\sim \xi(1-\xi)^n,\qquad n=9,11\\
\label{DA2} 
&&\phi_2(\xi)\sim \xi\exp(-m_b \xi/\beta_B).
\end{eqnarray}
The DA (\ref{DA2}) was proposed in \cite{neubertgrozin} in the context of heavy-quark effective theory. We use this parametrization for comparison 
although it does not strictly vanish at $\xi=1$. 

Fig.~\ref{Plot:4} compares the result corresponding to DA (\ref{DA1}) for $n=9$ and to (\ref{DA2}); for the latter, 
we fix the parameter $\beta_B$ such that $\langle{\xi^{-1}}\rangle$ for both parametrizations are equal to each other, 
$\langle{\xi^{-1}}\rangle=11$; 
this requirement yields $\beta=0.38$ GeV. Obviously, for a fixed value of $\langle{\xi^{-1}}\rangle$, 
the form factor is weakly sensitive to the precise shape of the $B$-meson DA: the sensitivity does not 
exceed a level of a few percent, at least in the region where our 
calculation of the form factors may be directly applied to the data. A similar picture is seen for other 
values of $q^2$ and $q'^2$. 

\begin{figure}[t!]
\begin{tabular}{cc}
\includegraphics[width=7cm]{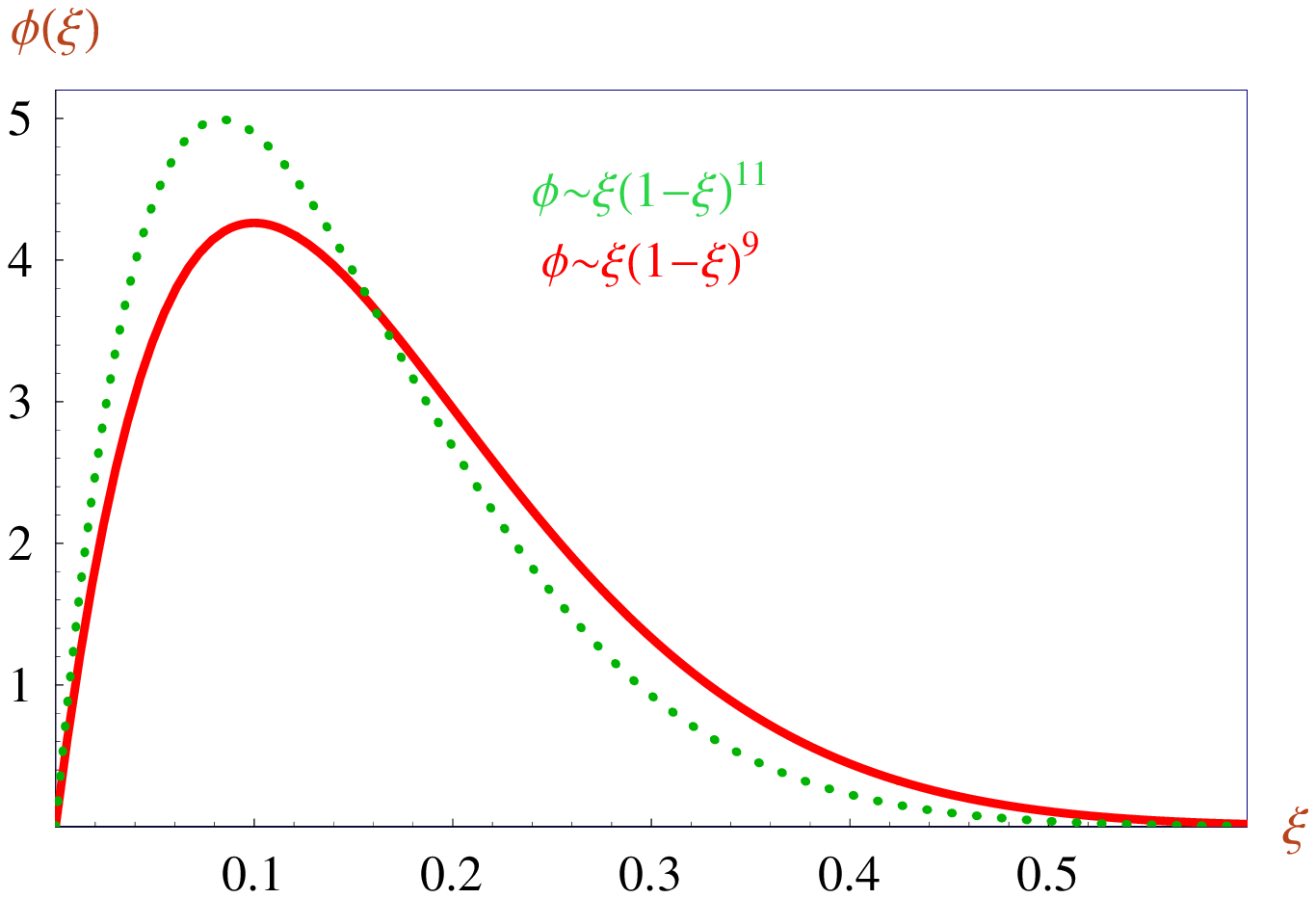} & \includegraphics[width=7cm]{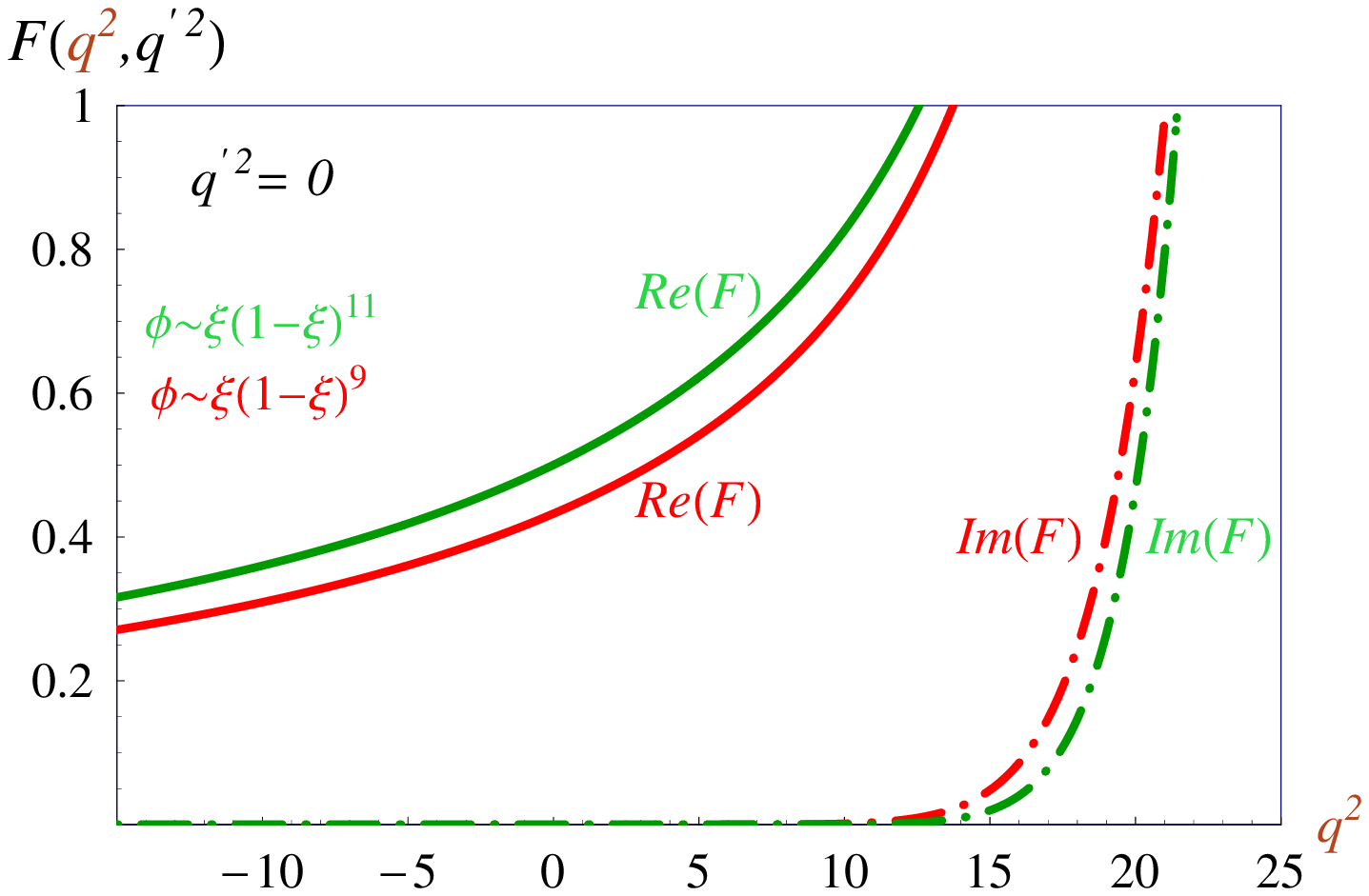}\\
(a) & (b)\\
\end{tabular}
\caption{\label{Plot:5}  
(a) The DA $\phi(\xi)\sim \xi(1-\xi)^n$ for $n=9$ with $\langle{\xi^{-1}}\rangle=11$ (red) 
and for $n=11$ with $\langle{\xi^{-1}}\rangle=13$ (green). (b) The corresponding $F(q^2,q'^2)$ vs $q^2$ for a fixed $q'^2=0$.}
\end{figure}

Fig.~\ref{Plot:5} illustrates the sensitivity of the form factor to the value of $\langle{\xi^{-1}}\rangle$ of the 
corresponding $B$-meson DA $\phi(\xi)$: we compare the form factor evaluated with 
DAs of (\ref{DA1}) for $n=9$ ($\langle{\xi^{-1}}\rangle=11$) and for $n=11$ ($\langle{\xi^{-1}}\rangle=13$). 
Indeed, as could be expected from the results of Section \ref{Sect:4}, 
there are regions of $q^2$ and $q'^2$, where the sensitivity to the value of $\langle{\xi^{-1}}\rangle$ of the 
$B$-meson DA is strong. 

\section{Discussion and Conclusions}
We presented the analysis of the analytic properties of the form factor $F(q^2,q'^2)$ which describes the $B$-decay into two currents.
A practically interesting application of this object is the decay of $B$-meson into four leptons. 

The results obtained above are based on the integral representation for the form factor $F(q^2,q'^2)$ in terms of 
the $B$-meson field-theoretic wave function and the quark propagator, for which the Feynman form is used. 
One knows, however, that the QCD confinement distorts this propagator in the soft region 
of momentum transfers: the Feynman propagator provides a good approximation 
to the full quark propagator in the region where the quark is highly virtual; in the problem under consideration it is sufficient 
that the quark virtuality is of order $\Lambda_{\rm QCD}m_b$. 
In practice, when one calculates the $\xi$-integral for the form factor in the $B$-decay kinematics, i.e. for $\sqrt{q^2}+\sqrt{q'^2}\le M_B$, 
one always picks up a contribution from the integration region where the quark propagator is not sufficiently hard. 
This is however not a great problem as soon as the major part of the form factor comes from the integration region where the quark is highly virtual. 
The region where the quark is soft leads to the appearance of the imaginary part of the form factor. We therefore conclude that 
{\it the results for the decay form factor obtained by calculating the correlation function (\ref{ff}) with 
Feynman quark and gluon propagators may be directly applied to the analysis of the realistic $B$-decay in those regions 
of $q^2$ and $q'^2$ where the imaginary part of the calculated form factor is much smaller than its real part.}  
This criterion may be used as a guiding principle for verifying the theoretical predictions for the form factors of interest and 
for restricting phenmenological parametrizations.

We studied the sensitivity of the form factor $F(q^2,q'^2)$ to the $B$-meson DA. 
It was found that the form factor is weakly sensitive to the precise shape of the $B$-meson DA, provided its 
first inverse moment $\langle{\xi^{-1}}\rangle$ is fixed.  
On the other hand, there are regions of $q^2$ and $q'^2$, where the form factor exhibits a sizeable 
sensitivity to the precise value of $\langle{\xi^{-1}}\rangle$ of the $B$-meson DA. 

Obviously, for obtaining the predictions for form factor of $B$-decays in QCD one should include quark spins into the game. 
Neverhteless, the presented analysis of the analytic properties of the form factors and the predictions for the behaviour 
of the form factor in the specific regions of $q^2$ and $q'^2$ are quite general and hold also in the real QCD. 

Some of these predictions are listed below: 
\begin{itemize}
\item[(i)]
In the region $q'^2=q^2$, and $|q^2|\gg M_B\Lambda_{\rm QCD}$, the leading behaviour of the form factor reads  
\begin{eqnarray}
F(q^2,q^2)=\frac{f_B}{-q^2}.
\end{eqnarray}
\item[(ii)]  
For $q'^2\simeq \Lambda_{\rm QCD}^2$, the form factor in a broad range of $q^2$ has a monopole behaviour  
\begin{eqnarray}
F(q^2,q'^2\simeq \Lambda_{\rm QCD}^2)= 
\frac{1}{(M_B^2-q^2)}\int\limits_0^1 d\xi \frac{\phi(\xi)}{\xi}.
\end{eqnarray}
\item[(iii)]  
In the region $M_B\Lambda_{\rm QCD} \ll q'^2\le M_B^2$ and for all $0 < q^2 < M_B^2$, the form factor has the form 
\begin{eqnarray}
F(0<q^2< M_B^2, q'^2\simeq M_B^2) = \frac{f_B}{-q'^2}\left(1+O(\Lambda_{\rm QCD}/m_b)\right).
\end{eqnarray}
\end{itemize}
The study presented here is just the first step in the analysis of complicated amplitudes of $B$-decays into two currents. 
We believe however that this is a useful step as it describes the general analytic properties of $F(q^2,q'^2)$ in quantum 
field theory, valid also for form factors in QCD. 

\vspace{-.10cm}
\acknowledgments{D.~M. gratefully acknowledges support by the joint RFBR/CNRS project 19-52-15022. 
S.~S. thanks the Italian Ministry of Research (MIUR) for support under the grant PRIN 20172LNEEZ. 
This work was done during $stay@home$ international action to stop the COVID-19 epidemic.}

\end{document}